# Using Topology to Predict Electrides in the Solid State

*Stefano Racioppi*, Eva Zurek*

Department of Chemistry, State University of New York at Buffalo, 777 Natural Science Complex, Buffalo, New York, 14260-3000, United States

KEYWORDS: Electride • Crystal Structure Prediction • Multi Objective • Electron Density • $Ca_5Pb_3$

**Abstract**

Electrides are characterized by electron density highly localized in interstitial sites, which do not coincide with the interatomic contacts. The rigorous quantum mechanical definition of electrides is based upon topological criteria derived from the electron density, and in particular the presence of non-nuclear attractors (NNAs). We employ these topological criteria in combination with crystal structure prediction methods (the XtalOpt evolutionary algorithm), to accelerate the discovery of crystalline electrides at ambient and non-ambient pressures. The localization and quantification of NNAs is used as the primary discriminator for the electride character of a solid within a multi-objective evolutionary structure search. We demonstrate the reliability of this approach through a comprehensive crystal structure prediction study of $Ca_5Pb_3$ at 20 GPa, a system previously theorized to exhibit electride character under compression. Our strategy could predict, and sort on-the-fly, several unknown low-enthalpy phases that possess NNAs in interstitial loci, such as the newly discovered *P4/mmm* structure. These results demonstrate how evolutionary algorithms, guided by rigorous topological descriptors, can be relied upon to effectively survey complex phases to find new electride candidates.

**Introduction**

Electrides are systems exhibiting highly localized electron density in interstitial voids. They have garnered interest from numerous scientific communities due to their intriguing electronic structure,[1–5] and because some of their properties are useful in industrial applications.[6–8] In the solid state, some electrides can have band-gaps comparable to electrical insulators or semiconductors,[9,10] while others can be metals,[1,11] superconductors,[12,13] ionic-conductors,[14] or even possess complex topological electronic structures.[15,16]

Electrides are typically recognized and characterized using topological methods,[17–20] such as the Quantum Theory of Atoms in Molecules (QTAIM)[21] and Electron Localization Functions (ELFs)[22]. On one hand, QTAIM studies properties of the electron density, $\rho(\mathbf{r})$, by mapping critical points and evaluating local characteristics. Since $\rho(\mathbf{r})$ is a quantum mechanical observable, QTAIM has become very popular in the field of quantum crystallography.[11,23] On the other hand, ELF highlights regions of space characterized by low Pauli repulsion, or in other words, Fermi holes.[24] The ELF takes values from 0 to 1, where 0.5 equals the value for a free electron gas spin probability distribution, and 1 corresponds to a zero probability of finding two electrons with the same spin at the same position (forbidden by Pauli exclusion). Values of ELF below 0.5 correspond to portions of space in which the electron density is more dilute than in a free electron gas, and therefore, it is the opposite of a localization region (See Ref. [20] for further information). Topologically, an electride must fulfill at least three criteria,[18,20] and possess: 1) a non-nuclear attractor (NNA – sometimes referred to in the literature as a non-nuclear maximum or NNM), a local maximum of the electron density that does not coincide with the position of a nucleus,[25,26]; 2) a negative value of the Laplacian of the electron density, $\nabla^2\rho(\mathbf{r})$, at the NNA position (which shows that $\rho(\mathbf{r})$ is a maximum at that local position); and 3) the presence of an ELF basin centered at the NNA, whose isosurface can still be seen with values as close to 1 as possible. This set of topological rules is generally valid for organic and inorganic electrides, at ambient and non-ambient conditions,[18,20] and is necessary to characterize electrides from a quantum mechanical point of view.

However, ambiguous cases exist. The notorious inorganic system $Ca_{12}Al_{14}O_{33}$ (Mayenite),[27] formed by a positive $[Ca_{12}Al_{14}O_{32}]^{2+}$ framework, in which free O atoms can be removed from the interstitials leaving behind an excess of free electrons to balance the charge,[27] has rapidly become the poster child of inorganic electrides. However, Density Functional Theory (DFT) calculations showed that the electron density inside the cavities is so diffuse that the compound lacks both NNAs and high ELF values.[17] Moreover, because of the high degree of delocalization of the conducting electrons, some studies have suggested that the "electride" label may be inappropriate for $Ca_{12}Al_{14}O_{32}$.[28] On the other hand, subsequent studies showed the presence of constructive overlap between Ca orbitals inside the cage of $[Ca_{12}Al_{14}O_{32}]^{2+}(e^-)_2$,[29,30] forming very diffuse multi-center bonds, which is a mechanism that has been proposed for the peculiar electron density characteristic of the electride state in high pressure inorganic electrides.[20] Further, it appeared that the possible electride state of $[Ca_{12}Al_{14}O_{32}]^{2+}(e^-)_2$ could be disordered in nature, causing its characteristic delocalized electron density,[28] and therefore, the lack of signatures like NNAs.[17,31]

Apart from such exceptional cases, it is clear that reliable descriptors are needed for developing efficient tools for the computational prediction of electride-like materials. In this context, the use of the ELF, in combination with the crystal structure prediction code CALYPSO, was proposed by Zhang and coworkers for the discovery of novel electrides.[32] In this pioneering study, the authors designed a methodology whereby they could remove regions within the unit cell that gave high ELF values along covalent bonds and associated with atomic cores, to find interstitial regions encapsulated by an ELF isovalue of 0.75. The ratio of the volume of this high-ELF region to the volume of the unit cell was employed as an electride-descriptor, and used as the searching criterion in the CALYPSO search. Though this strategy was shown to be effective, it suffers several drawbacks. Primarily, it requires the identification of several internal geometrical parameters, such as atomic radii and bond length distances, which are required to remove the values of the ELF that arise from anything that is not an interstitial "electride blob". These parameters were fixed, though in principle they should be adjusted for different conditions (*e.g.* pressure) and chemical stoichiometry. However, it is also now more recognized that the ELF alone is not a reliable descriptor for the electride state,[17–20] which has led to several misidentifications in the past years. Furthermore, this methodology only employed a single objective in the search and did not take into account energetic stability.

A different strategy for high-throughput identification of electrides was proposed by Burton and coworkers.[33] By counting the eigenstates near the Fermi level and evaluating the Bader charges within a range of 2.2 Å from the nuclei for almost 70,000 compounds deposited in a large database containing DFT data[34] they could identify 69 possible electrides, and retrieve some of those for which the electride state was already hypothesized. One drawback of this method is that it is a screening technique, but cannot be used to discover materials that are not already present within databases (in this case the Materials Project), precluding the possibility of discovering systems such as high pressure electrides. For similar reasons, the Mayenite system could not be identified as only the parent compound, $Ca_{12}Al_{14}O_{33}$, is a member of this database, and the electride requires oxygen deintercalation to produce the electride state. Moreover, since this method focuses only on materials characterized by many bands crossing the Fermi level, or in other words, metals, it would not be able to find insulating or semi-conducting electrides.

In the present work, we present our new method to accelerate the discovery of stable and metastable electrides in crystalline materials at ambient and non-ambient conditions. This methodology overcomes the aforementioned limitations in the previously proposed methodologies by combining fast topological analysis of the electron density with a multi-objective strategy to guide generative crystal structure predictions. We first present the technical aspects of our method, introducing the evolutionary algorithm-based crystal structure prediction code XtalOpt, the newly implemented multi-objectives strategy, and the QTAIM for the topological analysis of the electron density. Then, we show the results for the application of our method to a testcase, $Ca_5Pb_3$.

**Methodology**

**XtalOpt, an Evolutionary Algorithm-Based Code for Crystal Structure Prediction**

One successful crystal structure prediction (CSP) method for the discovery of hitherto unknown compounds employs genetic or evolutionary algorithms (GAs and EAs), which are optimization techniques based on concepts from biological evolution, such as selection, mutation, and reproduction.[35] XtalOpt, an open-source program for CSP, is among those codes designed to leverage this type of EA strategy.[36–38] Although known or targeted structures can also be purposefully introduced as "seeds", the EA process usually begins by creating an initial population of chemically reasonable random structures.[39] To find the nearest minimum on the potential energy surface (PES), each structure is subjected to local optimization using an external program. While in the past DFT codes were employed for local optimization, the field is moving in the direction of machine learning (ML) interatomic potentials,[40–42] which could dramatically accelerate such searches. Each structure's fitness has traditionally been gauged by the enthalpies or energies that are computed, and these values in turn affect the likelihood that a structure will be chosen as a parent for the construction of new structures.[38] Similar to genetic traits, those chemical systems with better qualities (in this case lower energy or enthalpy) pass on their structural characteristics to subsequent generations, gradually reshaping the population. One parent structure can be mutated to produce offspring structures, or features from two parent structures can be combined to create child structures. The evolutionary operations employed within XtalOpt within fixed composition[37] and variable composition[42] searches have been described in detail elsewhere. Briefly, to enable the algorithm to thoroughly explore nearby regions of the PES, mutations may entail minute modifications such as changing cell parameters, switching the locations of various atom types, or displacing atoms via random or periodic movements.[36,38] The real-space cut-and-splice method,[43] is a particularly powerful operation for joining two structures, encouraging more thorough investigation of the PES. Recently, XtalOpt has been extended to perform variable composition searches for multi-element systems, and introduced new operators that change the chemical composition of the system.[42]

However, selecting an offsprings only based on the relative energy, or enthalpy, of the predicted structures, might not be the ideal strategy when the goal of the CSP search is to find systems with specific characteristics, such as desired structural or electronic features, like in the case of electrides.[18] In fact, it is not necessarily true that structures with interesting features or properties are also the ground states of a given stoichiometry at certain pressure and temperature conditions. For example, diamond, one of the allotropes of carbon, is a superb superhard material, but it is metastable at ambient conditions. The need of focusing CSP searches towards other features, or objectives, has catalyzed the development of multi-objective evolutionary search methods.[38] As a testament of this initial effort, multi-objective strategies using XtalOpt have been put forward for the prediction of hard materials,[44] and methods that employ the similarity between a reference powder X-ray diffraction pattern and that calculated for an optimally-distorted structure generated in the EA search have been developed.[45]

**Multi-objective Search in XtalOpt**

Our proposed methodology for the prediction of electrides builds upon a multi-objective global optimization framework, as implemented in version 13.0 of the XtalOpt evolutionary algorithm.[38] In this version, the fitness of each predicted structure is evaluated based on multiple objectives defined by the user, which can include the enthalpy of the structure, alongside with other relevant physical, chemical or structural features. Following the standard geometry optimizations step, XtalOpt automatically invokes user-specified external programs to compute the desired objectives, which can be then combined with the enthalpy to determine the multi-objective fitness score. Another strategy for finding metastable systems with particular characteristics could be performed by filtering the gene pool,[46] also implemented in this version.

In the present study, two objectives were considered: the maximization of a topological descriptor of the electron density (denoted as $T$) and the minimization of the enthalpy ($H$). These objectives are incorporated into a single fitness function using a weighted linear combination:

$$f_s = w \left( \frac{T_s - T_{min}}{T_{max} - T_{min}} \right) + (1 - w) \left( \frac{H_{max} - H_s}{H_{max} - H_{min}} \right) \quad (1)$$

where $T_s$ and $H_s$ represent the values for the structure $s$. The parameters $A_{min}$ and $A_{max}$, with $A = T, H$, denote the minimum and maximum values of each objective within the current pool of structures. The weighting factor, $w$, defines the relative importance assigned to the objectives, ranging between 0 and 1. The sum of weights for all objectives is fixed to unity to properly balance their contributions to the total fitness, $f_s$. This fitness value is then used by XtalOpt to rank the structures generated by the CSP routine, guiding the selection of new parent structures for evolving the next generations.[36] In the just-released version of XtalOpt, version 14, Pareto optimization was implemented as an alternative to the sum of weights for all the objectives.[42] Two different strategies were implemented, both the standard technique where the Pareto front and crowding distances of the structures in the parent pool are employed to choose a parent using tournament selection, and another that calculates a Pareto-based fitness. Further details can be found in Ref. 42. If desired by the user, either one of these Pareto-based techniques could be employed as an alternative to the multi-objective strategy described here.

**Topological Analysis of the Electron Density in Electrides**

Our method exploits the topological analysis of the electron density, ρ(**r**), to automatically identify systems characterized by the presence of charge concentrations at special critical points named *non-nuclear attractors* (NNAs) and promote these quantum mechanical characteristics within the evolutionary search. The NNA is a topological descriptor that will take the role of the objective $T$ shown in Eq. 1. The topological analysis of the electron density is based on the Quantum Theory of Atoms in Molecules (QTAIM), invented by Richard Bader.[47] At the core of QTAIM is the mapping of critical points of the electron density, evaluated as

$$\nabla \rho(\boldsymbol{r}) = 0 \tag{2}$$

and the partition of ρ(**r**) into discrete, and well-defined, atomic entities.[48–50] The atomic boundaries are defined by the surface zero-flux condition of the electron density calculated for each atom in the system as[51]

$$\nabla \rho(r) \cdot \boldsymbol{n}(r) = 0 \text{ at every point on the surface} \tag{3}$$

where $\boldsymbol{n}(r)$ is the unit vector norm to the surface and perpendicular to $\nabla \rho(\mathbf{r})$, which is tangent to its gradient path at each point. The region of space defined by the zero-flux condition is called an atomic basin, a physically meaningful entity that satisfies the virial theorem.[52] Atomic basins have well-defined shapes (usually non-spherical) and volumes, and they can be used to integrate electronic charges and other properties.[53,54] Inside and at the surface of these basins exist critical points whose qualities are revealed by the topological analysis of ρ(**r**), and classified using two numbers, (ω, σ), corresponding to the number of non-zero curvatures (ω = rank of the critical point), and the sum of the algebraic signs of the Hessian eigenvalues (σ = signature of the critical point).[55] Therefore, in three dimensions, four critical points are possible:

(3, -3) (*non-*)*nuclear attractor* = all curvatures are negative, *i.e.*, a local maximum.

(3, -1) *bond critical point* = two curvatures are negative and one is positive.

(3, +1) *ring critical point* = two curvatures are positive and one is negative.

(3, +3) *cage critical point* = all curvatures are positive, *i.e.*, a local minimum.

Depending on the type of system (either a molecule in gas-phase or a periodic solid), the types and the numbers of critical points have to satisfy the following equations:

$$n - b + r - c = 1; \text{ in gas-phase systems (Poincaré-Hopf)} \qquad (4)$$

$$n - b + r - c = 0; \text{ in periodic systems (Morse)} \qquad (5)$$

Where $n, b, r, c$ refer to the four critical points listed above.[56,57]

An atomic nucleus is a nuclear attractor of the electron density, and its position is marked by the presence of a maximum in ρ(**r**), a (3,-3) critical point. However, Richard Bader and Carlo Gatti, were the first to notice, in the late 1980s, that some alkali metals clusters are characterized by a very peculiar feature of the electron density: a maximum in ρ(**r**) without a nucleus,[26,58] which was therefore named a non-nuclear attractor, NNA. Being a maximum of the electron density, an NNA also possesses a topological basin. It was then recognized that the presence of NNAs is a main topological feature in electrides,[18] and generated as a consequence of the formation of multi-center bonds (MCBs) (Figure 1).[1,19,59]

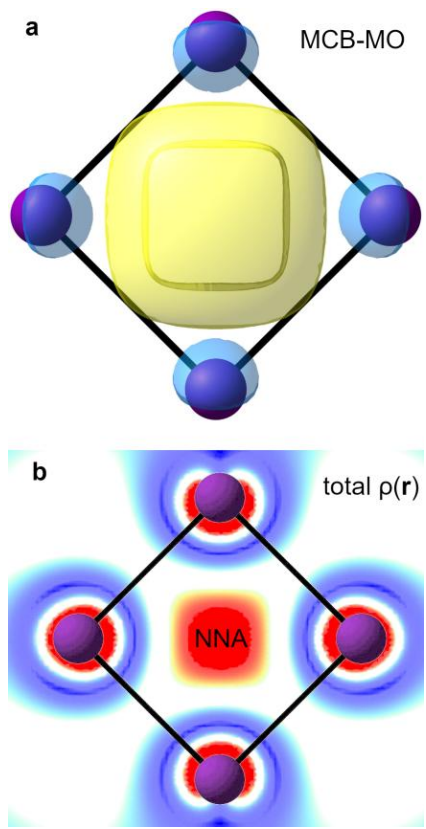

**Figure 1.** Na$_4$ cluster, calculated in the gas-phase with Gassian16[60] at the HSE06/aug-cc-PVQZ level of theory, where we rendered: **a**) The multi-centered bond (MCB) molecular orbital (MO) in the Na$_4$ cluster (isosurface 0.06). **b**) The total electron density, $\rho(\mathbf{r})$, of Na$_4$ passing through the molecular plane. The red areas correspond to high electron density values, while the blue areas are values close to zero (range [0.0 – 0.015]). The inner red region is where the NNA lies.

Due to their rigorous definition and unique characteristics, we decided to exploit NNAs in our multi-objective evolutionary search. In this work, the level of "electrideness" of a system, is quantified from the value of the electron density at the NNA, $\rho_{NNA}$ [e/Å$^3$] ($T$ in Eq. 1), which will be then used as a component in the calculation of the fitness in the CSP. Operatively, the process checks for the presence of an NNA and then quantifies $\rho_{NNA}$, which becomes the value assigned to the objective. If no NNA is found, a value of zero is assigned to the objective, but the structure is still kept, since the enthalpy is also used as secondary objective. However, one could also use a filtration operation to discard all the structures that do not possess an NNA.[46] To be noted, other topological descriptors could have been used in combination with the electron density at the NNA,[11] such as the value of the Laplacian, $\nabla^2\rho_{NNM}$ [e/Å$^5$], which quantifies the curvature, *i.e.*, the degree of concentration of $\rho(\mathbf{r})$ at the critical point; or even the value of the ELF,[61] which in electrides should have values close to 1.[20] Additionally, instead of using local properties of the electron density, like $\rho_{NNA}$ and $\nabla^2\rho_{NNM}$, or the value of the ELF at the NNA, one could set an objective based on an integrated property. As we described before, to the NNA is associated a basin (Eq. 3), which could be used to integrate part of a scalar field. For example, integrating the electron density using the volume of the NNA basin, one would obtain the total charge, $q$, associated with the region of space of the electride as

$$\int_{V_{NNA}} \rho(\mathbf{r})\, d\mathbf{r} = q \tag{6}$$

Using Eq. 6 to estimate the level of electrideness of a crystalline material is probably a very valid choice. However, the drawback of using this quantity as an objective in a CSP search is that accurate integrations are usually computationally demanding, and therefore slower than the simple location and quantification of critical points. In fact, a complete critical point search usually takes only a few seconds on a desktop computer,[57] and therefore, it does not add any significant computational cost to the structure search. On the other hand, the integration over basins can take several minutes or more, depending on the size of the grid that describes the scalar field, with a significant computational impact when thousands of structures are examined.

By using $\rho_{NNA}$ in our multi-objective search as a topological descriptor for electrides, we expect that the evolutionary algorithm will favor those systems that possess structural features that promote the generation of NNAs in interstitial loci, which will be promoted as parents in the next generation. The quantification of $\rho_{NNA}$ is automatically done by the code Critic2[62] for each system during the crystal structure search, and it does not require any type of parametrization.

**Computational Details**

The open-source evolutionary algorithm XtalOpt[36,38] version 13.0 was employed for the CSP search using multi-objective global optimization. We also performed a CSP search using the newly-implemented default Pareto front settings as a test,[42] but did not observe any significant improvement or difference compared to the originally proposed multi-objective global optimization strategy.[38] The parameters employed in XtalOpt are thoroughly described in Ref.37, and the weighting factor for the electride-objective (value of $\rho_{NNA}$) was set to 0.7. This choice also promotes the choice of parents with low energies or enthalpies, unlike previous work performed using the CALYPSO methodology, which could have, in principle, located electrides that are too high in energy to be synthesizable.[63] The initial generation consisted of 200 random symmetric structures that were created by the RandSpg algorithm,[39] while the parents pool size was set equal to 20. The number of formula units for the $Ca_5Pb_3$ stoichiometry was set from 1 to 4. A sum of the atomic radii scaled by a factor of 0.7 was used to determine the shortest distances allowed between pairs of atoms. Duplicate structures were identified and removed from the breeding pool using the XtalComp algorithm.[64] The total number of generated structures was equal to 1000. Each structure search followed a multi-step strategy, with three subsequent optimizations with increasing level of accuracy, plus a final accurate single point, which was followed by the topological analysis of the structure's electron density.

Geometry optimizations and electronic structure calculations were performed using Density Functional Theory (DFT) with the Vienna Ab Initio Simulation Package (VASP), version 6.4.2.[65] The PBE[66] exchange-correlation functional was employed. The projector augmented wave (PAW) method[67] was used to treat the core states in combination with a plane-wave basis set with an energy cutoff of 300 eV for the geometry optimizations and 600 eV for the final single point. The Ca $3p^64s^2$ (PAW_PBE Ca_pv) and Pb $6s^26p^2$ (PAW_PBE Pb) states were treated explicitly for the geometry optimizations, while Ca $3s^23p^64s^2$ (PAW_PBE Ca_sv) and Pb $5d^{10}6s^26p^2$ (PAW_PBE Pb_d) states were treated explicitly for the final single point. The non-spherical contributions related to the gradient of the density in the PAW spheres were included. The k-point meshes were generated using the Γ-centered Monkhorst−Pack scheme,[68] and the number of divisions along each reciprocal lattice vector was selected so that the product of this number with the real lattice constant was greater than or equal to a given value. The values of 30, 40 and 50 Å were used for the three subsequent optimization steps in the crystal structure search, then a value of 60 Å was used for the final single point. The accuracy of the energy convergence was set to increase from $10^{-3}$ to $10^{-5}$ eV for the optimizations, and to $10^{-6}$ eV for the final single point on the structures for which the norms of all the forces calculated during the relaxations were smaller than $10^{-3}$ eV/Å. A Gaussian smearing was used at each optimization step, and for each system with a sigma of 0.02 eV. The tetrahedron method was adopted in the last single point.[69]

For the projected density of states, the Wigner-Seitz radius of Ca and Pb were re-evaluated at high pressure based on the QTAIM basins' volumes calculated with the Critic2 code,[62] and slightly increased to cover the entire volume of the unit cell (*i.e.*, to account for the NNA volume and perform a projection of all the states). Phonons in the harmonic approximation were determined for $Ca_5Pb_3$ *P4/mmm* with the Phonopy package[70] using a 2x2x3 supercell, containing 96 atoms and having lattice parameters greater than 10 Å. The crystal structure prediction was performed at a pressure of 20 GPa. The quantum chemical code Gassian16[60] was used to calculate the electron density of the $Na_4$ cluster at the HSE06/aug-cc-PVQZ level of theory shown in Figure 1.

**Results and Discussions**

We will use our new method for the prediction of crystalline electrides with the $Ca_5Pb_3$ stoichiometry under mild compression. The choice for this specific elemental composition stems from the fact that the *P6$_3$/mcm* phase of $Ca_5Pb_3$[71] is known, possessing a $Mn_5Si_3$-type of structure. This particular phase (*P6$_3$/mcm*) was calculated to be a high lying metastable system at ambient pressure conditions, lying 127 meV/atom above the convex hull, which suggest it could decompose into $Ca_2Pb$ and $CaPb$.[72] Yet, $Ca_5Pb_3$ was synthesized at ~ 1300 K at ambient pressure.[73] A recent theoretical study[72] proposed that $Ca_5Pb_3$ would remain in the *P6$_3$/mcm* phase upon compression to 20 GPa, while transitioning into an electride state. However, if other stable phases of $Ca_5Pb_3$ exist upon compression, and whether any of them is potentially an electride, is unknown. An evolutionary search identified a *Pm* symmetry $Ca_5Pb_3$ phase that was nearly isoenthalpic (0.5 meV/atom more stable) at 1 atm, but from the analysis of its electronic structure, it did not appear to be an electride. Therefore, we applied our newly developed methodology to this stoichiometry to discover new potential electrides.

In Figure 2a, we plot the calculated values of the electron density at the main NNA, $\rho_{NNA}$, for each unique predicted structure at 20 GPa, together with its enthalpic stability relative to the $P6_3/mcm$ phase.[72] In this context, the "main NNA" is the one with the highest $\rho_{NNA}$ value, in case more NNAs where found in the same structure.[12] The first result that appears immediately evident is that the $P6_3/mcm$ phase is neither the thermodynamic ground state at this pressure (20 GPa), nor is it the "most electride" phase (based on the value of $\rho_{NNA}$). Over the whole CSP prediction, which generated 1000 structures guided by the multi-objective search, 649 were unique phases,[64] and 368 of these possessed at least one NNA. Surprisingly, $P6_3/mcm$ has the second lowest value of $\rho_{NNA}$ among the 368 possible electride structures found by our search. Moreover, in Figure 2b, we also show the percentage of unique structures, at each generation, for which the presence of an NNA was confirmed upon performing the critical points analysis, versus the total number of unique structures in that generation. This result (Figure 2b) shows that the multi-objective search steers the CSP favoring those phases having structural features that promote the formation of NNAs.

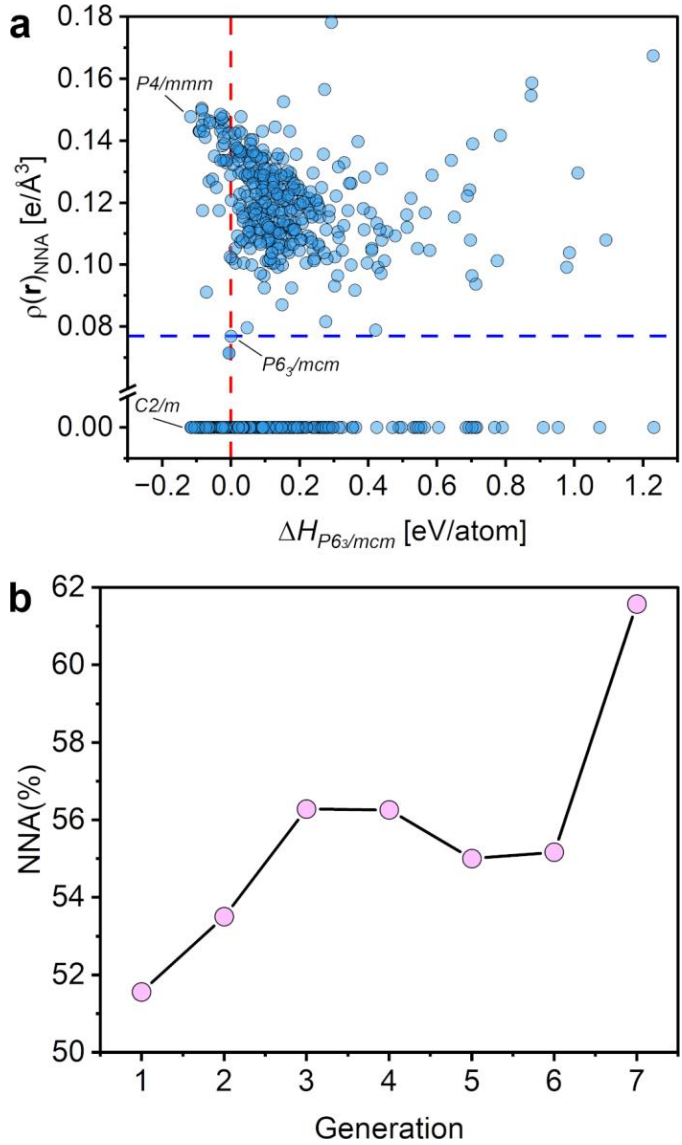

**Figure 2.** (**a**) Value of the electron density at the NNA position, $\rho_{NNA}$ [e/Å$^3$], per each unique Ca$_5$Pb$_3$ structure found in the CSP run performed at 20 GPa versus the stability, $\Delta H$ [eV/atom], relative to the $P6_3/mcm$ phase.[72] The blue and red dashed lines mark the values of $\rho_{NNA}$ and $\Delta H$ relative to the $P6_3/mcm$ phase. Values of $\rho_{NNA} = 0$ correspond to structures without NNAs. (**b**) The percentage of unique structures having an NNA over the total number of unique structures per generation. The first generation is completely random (200 structures), while from the second generation on the CSP is driven by the EA and the multi-objective fitness parameter.

Among the new structures having the highest $\rho_{NNA}$ values, the most stable found by our search has the *P4/mmm* space group, and it is ~120 meV/atom more stable than the previously proposed *P6₃/mcm* phase (Figure 3). An additional new structure having space group *C2/m*, and only a few meV/atom less stable than *P4/mmm*, was found by our search. Since no NNA was located in the *C2/m* phase upon topological investigation (corresponding to $\rho_{NNA} = 0$ in Figure 2a), it will not be further discussed.

The *P4/mmm* phase is dynamically stable at 20 GPa (Figure S1), and it is predicted to become thermodynamically preferred over *P6₃/mcm* at 6.5 GPa (Figure S2). The projected density of states (pDOSs) calculated with the meta-GGA PBE functional predict that these phases – the newly predicted *P4/mmm* (and *C2/m*), and the known *P6₃/mcm* phase – are metals, with the bands near the Fermi level mostly occupied by Ca 3d-orbitals and Pb 6p-orbitals (Figure S3). However, only *P4/mmm* possesses a pseudo-gap near the Fermi level characteristic of many known electride phases.[11,12]

As we mentioned in the introduction, the recognition of an electride requires the evaluation of several topological descriptors: the presence of an NNA in the electron density, large and negative Laplacian at the NNA critical points, and a high ELF value (close to 1).[18,20] The *P4/mmm* phase has a value of $\rho_{NNA}$ equal to 0.15 e/Å³, which is among the highest values found by our search (Figure 2a), and almost double compared to the one calculated for the NNA in *P6₃/mcm*. Further analyses reveals also that the Laplacian of the electron density, $\nabla^2\rho_{NNA}$, and the charge integrated at the NNA basin, *q*, are much larger for *P4/mmm* compared to the *P6₃/mcm* phase. Moreover, only the tetragonal phase shows an ELF value > 0.9 (Figure 3). Therefore, the combination of all these topological descriptors confirms the high level of "electrideness" of *P4/mmm* Ca₅Pb₃.

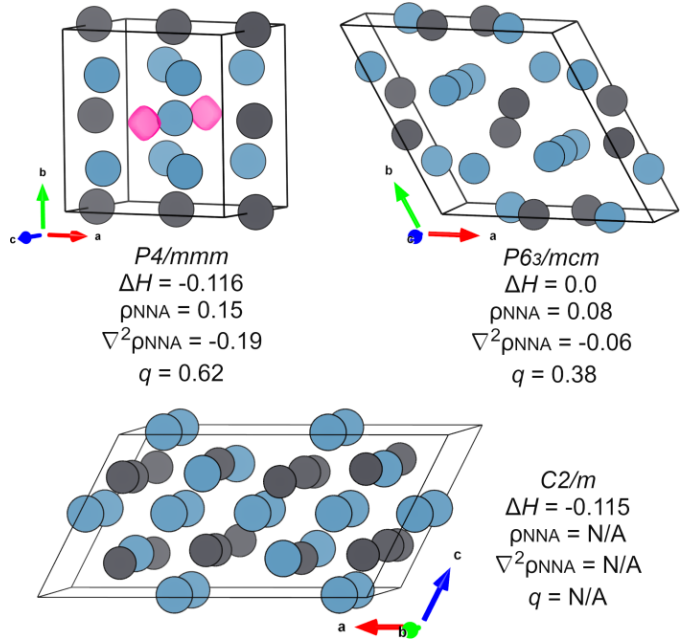

**Figure 3.** The most relevant structures predicted at 20 GPa for the Ca$_5$Pb$_3$ stoichiometry (color code: calcium = blue; lead = grey), reporting their enthalpies, $\Delta H$ in eV/atom, relative to *P6$_3$/mcm*.[72] The calculated value of the electron density and it's Laplacian at the NNA position, $\rho_{NNA}$ [e/Å$^3$] and $\nabla^2\rho_{NNA}$ [e/Å$^5$], and integrated charge, $q$, are also reported (if an NNA was located). The ELF isosurface for values > 0.9 is plotted in pink (found only in *P4/mmm*).

From the structural point of view, *P4/mmm*, *P6$_3$/mcm*, and other electride phases that were identified in the search, have something in common that appears to be crucial for the generation of NNAs: All possess some type of Ca-clusters within which the NNA is centered (Figure 4). In the cases of *P4/mmm* and *P6$_3$/mcm*, the clusters are Ca$_6$-octahedra. However, other types of clusters were also found in other structures. For example, many phases presented Ca$_4$-tetrahedral clusters, while the phase having the highest value of $\rho_{NNA}$ found by our CSP run (equal to 0.18

e/Å$^3$ and $\Delta H = 0.29$ eV/atom, see Figure 2a), possessed a square-based pyramid Ca$_5$-cluster (Figure 4). Therefore, since the multi-objective search promotes phases possessing NNAs, it indirectly promotes also structures characterized by Ca-clusters, since NNAs are located at their centeres. In other words, the simultaneous presence of NNAs and Ca-clusters, which is an intrinsic structure-to-property relationship in Ca$_5$Pb$_3$, has been exploited by the topological multi-objective CSP search to predict new electride phases. Similar relationships have previously been observed in calculations performed on Li[12] and Ca[11] phases under compression, as well as on Li$_n$ aggregates in the gas-phase,[26] specifically that NNAs were always located within clusters formed by the metallic element.

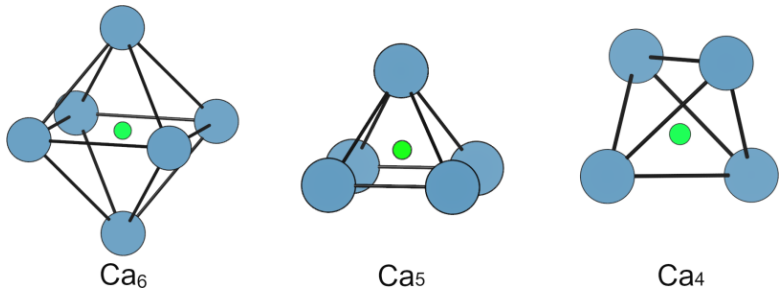

**Figure 4.** Details of some calcium clusters surrounding the NNAs (green dots) present in different electride phases of Ca$_5$Pb$_3$ found in our multi-objective crystal structure search.

**Conclusions**

In this work, we have introduced a quantum mechanically robust, parameter-free, approach towards electride identification based on the topology of the electron density, and presented its incorporation in a multi-objective evolutionary crystal structure prediction framework. Through the location of non-nuclear attractors (NNAs) as a direct and unambiguous criterion for electride character, we have identified a link between predicted crystal structures and their electronic character. Our multi-objective strategy implemented in XtalOpt, explored the $Ca_5Pb_3$ system under pressure and identified several novel low-enthalpy phases, including a hitherto unreported, and dynamically stable, *P4/mmm* structure. This work demonstrates how directly integrating electron density descriptors into the prediction of structures can further enable the discovery of materials with desired properties, in particular in complex systems where such properties are sensitive to pressure-induced electronic reorganizations. The strategy that we outline is generic and can be applied for the discovery of electrides at any pressure conditions, offering an experimentally useful and tractable route to mapping the structure-property landscape of crystalline electrides. It overcomes the limitations of previous computational proposals for the identification of electrides as it (1) is not sensitive to the choice of tabulated parameters such as bond distances and atomic radii, (2) can simultaneously optimize electrideness and enthalpy, and (3) can generate hitherto unknown structures that are not present within existing databases.

ASSOCIATED CONTENT

**Supporting Information**. Crystal structures of the discussed $Ca_5Pb_3$ phases were uploaded as cif files. The electronic structure analysis and phonon results are reported in the Supporting Information.

AUTHOR INFORMATION

**Corresponding Author**

**Stefano Racioppi** – Department of Chemistry, State University of New York at Buffalo, 777 Natural Science Complex, Buffalo, New York, 14260-3000, United States; https://orcid.org/0000-0002-4174-1732

E-mail: sraciopp@buffalo.edu

**Eva Zurek** – Department of Chemistry, State University of New York at Buffalo, 777 Natural Science Complex, Buffalo, New York, 14260-3000, United States; https://orcid.org/0000-0003-0738-867X

E-mail: ezurek@buffalo.edu

**Author Contributions**

E.Z and S.R. conceived the project, and S.R. performed the calculations and analysis. The manuscript was written through contributions of all authors. All authors have given approval to the final version of the manuscript.

**Acknowledgments**

Funding for this research is provided by the Center for Matter at Atomic Pressures (CMAP), a National Science Foundation (NSF) Physics Frontier Center, under Award PHY-2020249 and the NSF award DMR-2119065. Calculations were performed at the Center for Computational Research at SUNY Buffalo (http:///hdl.handle.net/10477/79221)

ABBREVIATIONS

CSP: Crystal Structure Prediction.

DFT: Density Functional Theory.

ELF: Electron Localization Function.

MCB: Multi-center bond.

NNA: Non-nuclear attractor.

QTAIM: Quantum Theory of Atoms in Molecules.

# Supporting Information

# Using Topology to Predict Electrides in the Solid State

*Stefano Racioppi*, Eva Zurek**

Department of Chemistry, State University of New York at Buffalo, 777 Natural Science Complex, Buffalo, New York, 14260-3000, United States

**Phonons and Electronic Structure of *P4/mmm* at 20 GPa**

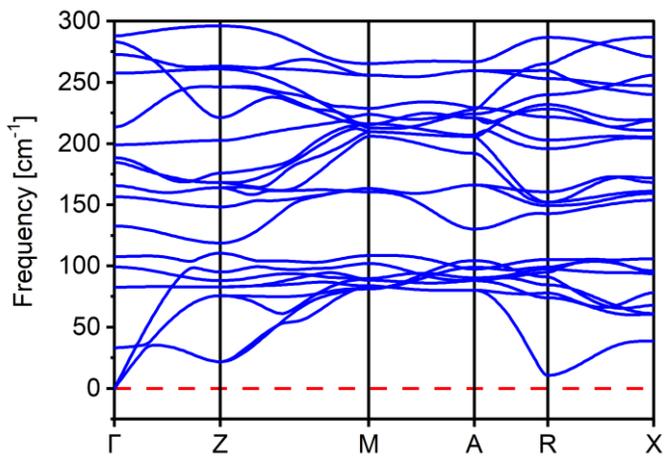

**Figure S1.** Phonon band structure of the optimized *P4/mmm* phase of $Ca_5Pb_3$ at 20 GPa, calculated with the VASP code using the PBE functional.

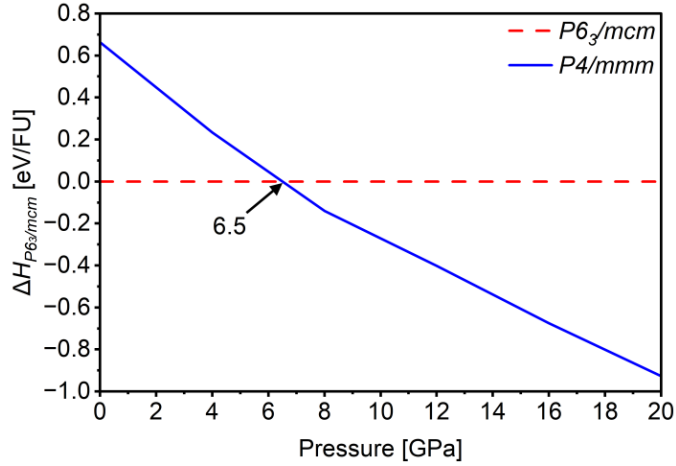

**Figure S2.** Enthalpy differences ($\Delta H = \Delta E + P\Delta V$) as a function of pressure and relative to the $P6_3/mcm$ phase of Ca$_5$Pb$_3$, calculated with the VASP code using the PBE functional.

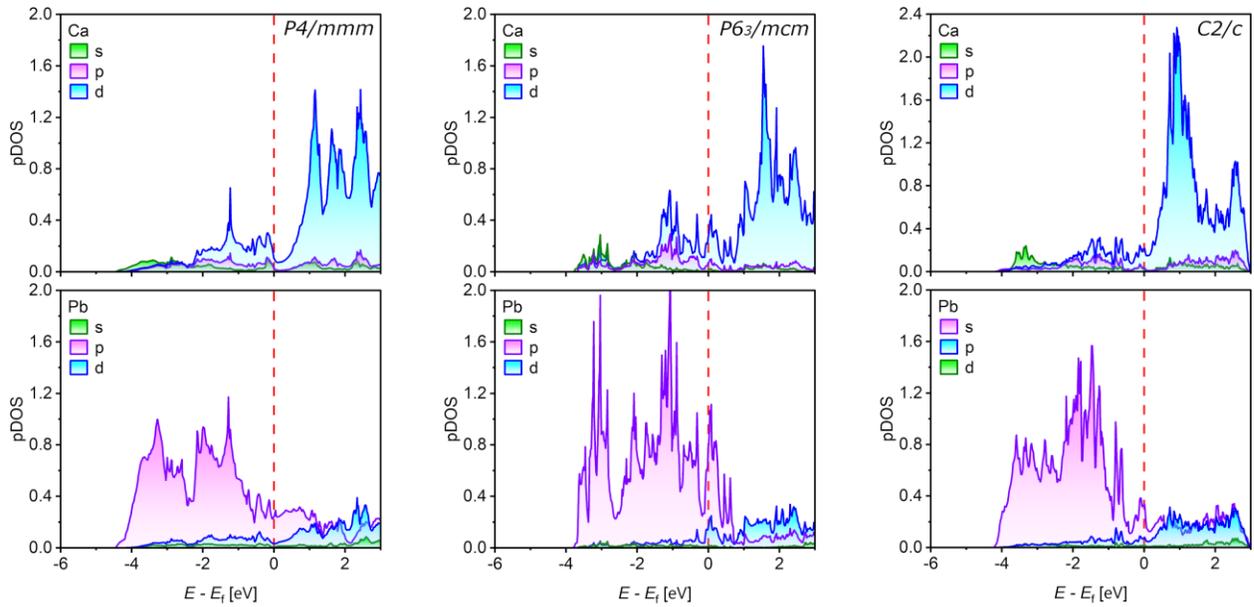

**Figure S3.** Orbital projected density of states (pDOS) per atom of the $P4/mmm$, $P6_3/mcm$ and $C2/c$ phases of Ca$_5$Pb$_3$ optimized at 20 GPa, calculated with the VASP code using the PBE functional. The top of the valence band is set to 0 eV.